\documentclass[twocolumn]{aastex63}

\usepackage{mathtools}

\newcommand{\hi}{H\,{\sc i}}
\newcommand\kms{km$\,$s$^{-1}$}

\usepackage[multiple]{footmisc}

\received{}
\revised{}
\accepted{}
\shorttitle{A Quenched, Isolated Ultra-Faint Dwarf Galaxy}
\shortauthors{Sand et al.}

\graphicspath{{./}{figures/}}

\begin{document}

\title{Tucana B: A Potentially Isolated and Quenched Ultra-faint Dwarf Galaxy at D$\approx$1.4 Mpc\footnote{This paper includes data gathered with the 6.5~m Magellan Telescope at Las Campanas Observatory, Chile.}}

\correspondingauthor{D. J. Sand}
\email{dsand@arizona.edu}

\newcommand{\Thacher}{\affiliation{Thacher Observatory, Thacher School, 5025 Thacher Rd. Ojai, CA 93023, USA}}
\newcommand{\UA}{\affiliation{Steward Observatory, University of Arizona, 933 North Cherry Avenue, Tucson, AZ 85721-0065, USA}}
\newcommand{\NU}{\affiliation{Center for Interdisciplinary Exploration and Research in Astrophysics and Department of Physics and Astronomy, \\ Northwestern University, 2145 Sheridan Road, Evanston, IL 60208-3112, USA}}
\newcommand{\UCDavis}{\affiliation{Department of Physics, University of California, 1 Shields Avenue, Davis, CA 95616-5270, USA}}
\newcommand{\Padova}{\affiliation{Department of Physics and Astronomy Galileo Galilei, University of Padova, Vicolo dell'Osservatorio, 3, I-35122 Padova, Italy}}
\newcommand{\INAF}{\affiliation{INAF Osservatorio Astronomico di Padova, Vicolo dell'Osservatorio 5, I-35122 Padova, Italy}}
\newcommand{\INAFbol}{\affiliation{INAF - Osservatorio di Astrofisica e Scienza dello Spazio - Via Piero Gobetti 93/3, I-40129 Bologna, Italy}}
\newcommand{\LPL}{\affiliation{Lunar and Planetary Lab, Department of Planetary Sciences, University of Arizona, Tucson, AZ 85721, USA}}
\newcommand{\NOAO}{\affiliation{National Optical Astronomy Observatory, 950 North Cherry Avenue, Tucson, AZ 85719, USA}}
\newcommand{\PATAU}{\affiliation{The School of Physics and Astronomy, Tel Aviv University, Tel Aviv 69978, Israel}}
\newcommand{\TTU}{\affiliation{Department of Physics and Astronomy, Texas Tech University, Box 1051, Lubbock, TX 79409-1051, USA}}
\newcommand{\OSU}{\affiliation{Department  of  Astronomy,  The  Ohio  State University,  140  W.  18th  Ave.,  Columbus,  OH43210, USA}}
\newcommand{\LBT}{\affiliation{Large Binocular Telescope Observatory, 933 North Cherry Avenue, Tucson, AZ, USA}}
\newcommand{\RMC}{\affiliation{Department of Physics and Space Science Royal Military College of Canada P.O. Box 17000, Station Forces Kingston, ON K7K 7B4, Canada}}
\newcommand{\ASU}{\affiliation{School of Earth and Space Exploration, Arizona State University, Tempe, AZ 85287, USA}}
\newcommand{\MMT}{\affiliation{MMT Observatory, PO Box 210065, University of Arizona, Tucson, AZ 85721-0065, USA}}
\newcommand{\NAU}{\affiliation{Department of Physics and Astronomy, Northern Arizona University, P.O. Box 6010, Flagstaff, AZ 86011, USA}}
\newcommand{\UAOptSci}{\affiliation{College of Optical Sciences, University of Arizona, 1630 E University Blvd, Tucson, AZ 85719, USA}}
\newcommand{\UNC}{\affiliation{Department of Physics and Astronomy, University of North Carolina at Chapel Hill, Chapel Hill, NC 27599, USA}}
\newcommand{\MSU}{\affiliation{Center for Data Intensive and Time Domain Astronomy, Department  of  Physics  and  Astronomy,  Michigan  State  University,East Lansing, MI 48824, USA}}
\newcommand{\UCSC}{\affiliation{Department of Astronomy and Astrophysics, University of California, Santa Cruz, CA 95064, USA}}
\newcommand{\STScI}{\affiliation{Space Telescope Science Institute, 3700 San Martin Drive, Baltimore, MD 21218, USA}}
\newcommand{\Brandeis}{\affiliation{Department of Physics, Brandeis University, Waltham, MA 02453, USA}}
\newcommand{\LCO}{\affiliation{Las Cumbres Observatory, 6740 Cortona Drive, Suite 102, Goleta, CA 93117-5575, USA}}
\newcommand{\UToronto}{\affiliation{Department of Astronomy and Astrophysics, University of Toronto, 50 St. George Street, Toronto, Ontario, M5S 3H4 Canada}}
\newcommand{\NotreDame}{\affiliation{Department of Physics, University of Notre Dame, Notre Dame, IN 46556, USA}}
\newcommand{\UMN}{\affiliation{College of Science \& Engineering, Minnesota Institute for Astrophysics, University of Minnesota, 115 Union St. SE, Minneapolis, MN 55455, USA}}
\newcommand{\UT}{\affiliation{Department of Astronomy, University of Texas at Austin, Austin, TX 78712, USA}}
\newcommand{\JHU}{\affiliation{The Johns Hopkins University, Baltimore, MD 21218, USA}}
\newcommand{\VAT}{\affiliation{Vatican Observatory, 00120 Citt\`{a} del Vaticano, Vatican City State  }}
\newcommand{\HF}{\affiliation{Hubble Fellow}}
\newcommand{\Carnegie}{\affiliation{The Observatories of the Carnegie Institution for Science, 813 Santa Barbara St., Pasadena, CA 91101, USA}}

\author[0000-0003-4102-380X]{David~J. Sand}
\UA

\author[0000-0001-9649-4815]{\textsc{Bur\c{c}{\rlap{\.}\i}n Mutlu-Pakd{\rlap{\.}\i}l}}
\affil{Kavli Institute for Cosmological Physics, University of Chicago, Chicago, IL 60637, USA}
\affil{Department of Astronomy and Astrophysics, University of Chicago, Chicago IL 60637, USA}
\affil{Department of Physics and Astronomy, Dartmouth College, 6127 Wilder Laboratory, Hanover, NH 03755, USA}

\author[0000-0002-5434-4904]{Michael G. Jones}
\UA
%\affiliation{Steward Observatory, University of Arizona, 933 North Cherry Avenue, Rm. N204, Tucson, AZ 85721-0065, USA}

\author[0000-0001-8855-3635]{Ananthan Karunakaran}
\affiliation{Instituto de Astrof\'{i}sica de Andaluc\'{i}a (CSIC), Glorieta de la Astronom\'{i}a, 18008 Granada, Spain}

\author[0000-0002-7633-431X]{Feige Wang}\thanks{NASA Hubble Fellow}
\UA

\author[0000-0001-5287-4242]{Jinyi Yang}\thanks{Strittmatter Fellow}
\UA

\author[0000-0002-7155-679X]{Anirudh Chiti}
\affil{Kavli Institute for Cosmological Physics, University of Chicago, Chicago, IL 60637, USA}
\affil{Department of Astronomy and Astrophysics, University of Chicago, Chicago IL 60637, USA}

\author[0000-0001-8354-7279]{Paul Bennet}
\affiliation{Space Telescope Science Institute, 3700 San Martin Drive, Baltimore, MD 21218, USA}

\author[0000-0002-1763-4128]{Denija Crnojevi\'{c}}
\affil{University of Tampa, 401 West Kennedy Boulevard, Tampa, FL 33606, USA}

\author[0000-0002-0956-7949]{Kristine Spekkens}
\affiliation{Department of Physics and Space Science, Royal Military College of Canada,\\ P.O. Box 17000, Station Forces Kingston, ON K7K 7B4, Canada}
\affiliation{Department of Physics, Engineering Physics and Astronomy, Queen’s University, Kingston, ON K7L 3N6, Canada}

\begin{abstract}
We report the discovery of Tucana~B, an isolated ultra-faint dwarf galaxy at a distance of D=1.4 Mpc.  Tucana~B was found during a search for ultra-faint satellite companions to the known dwarfs in the outskirts of the Local Group, although its sky position and distance indicate the nearest galaxy to be  $\sim$500 kpc distant. Deep ground-based imaging resolves Tucana B into stars, and it displays a sparse red giant branch consistent with an old, metal poor stellar population analogous to that seen in the ultra-faint dwarf galaxies of the Milky Way, albeit at fainter apparent magnitudes.  Tucana~B has a half-light radius of 80$\pm$40 pc, and an absolute magnitude of $M_V$=$-$6.9$^{+0.5}_{-0.6}$ mag ($L_V$=$(5^{+4}_{-2})\times$10$^4$ $L_{\odot}$), which is again comparable to the Milky Way's ultra-faint satellites.  There is no evidence for a population of young stars, either in the optical color magnitude diagram or in {\it GALEX} archival ultraviolet imaging, with the {\it GALEX} data indicating $\log (\mathrm{SFR_{NUV}/M_\odot \, yr^{-1}}) < -5.4$ for star formation on $\lesssim$100 Myr time scales.  Given its isolation and physical properties, Tucana B may be a definitive example of an ultra-faint dwarf that has been quenched by reionization, providing strong confirmation of a key driver of galaxy formation and evolution at the lowest mass scales. It also signals a new era of ultra-faint dwarf galaxy discovery at the extreme edges of the Local Group.
\end{abstract}


\section{Introduction} \label{sec:intro}

The faint end of the galaxy luminosity function is important for understanding dark matter and astrophysics on small scales \citep[see, e.g.][for recent reviews]{Bullock17,Simon19}.  In the Local Group, observations continue to find a variety of ultra-faint galaxies \citep[for instance, most recently][]{Mau20,Cerny21,Cerny22}, while numerical simulations work out how stars form in the smallest dark matter subhalos of Milky Way-like systems \citep[e.g.,][]{Brooks13,Sawala16,Wetzel16,Samuel20,Engler21,Applebaum21}. Outside of the Local Group, faint dwarf galaxies are being identified in resolved stars \citep{Chiboucas13,Sand14,Crnojevic14,Crnojevic16,Crnojevic19,Toloba16,Carlin16,Smercina18,Bennet19,Bennet20,Mutlu22}, diffuse or semi-resolved light \citep[e.g.][]{Bennet17,Carlsten20,Davis21}, as well as spectroscopic surveys \citep{Geha17,Mao21}.  These programs are leading to a new understanding of both the scatter in satellite properties as well as potential challenges to our picture of galaxy formation on small scales \citep{Bennet19,Bennet20,Carlsten21,Smercina21,Karunakaran21}.

Despite this progress, there are still regions of dwarf galaxy discovery space that are largely unexplored.   In particular, only a handful of new dwarfs have been uncovered at the periphery of the Local Group and its immediate environs ($D$$\approx$0.5--2.0 Mpc).  Examples include the star forming and relatively isolated dwarf Leo P \citep[D=1.6 Mpc, $M_V$=$-$9.3 mag; ][]{Giovanelli13,Rhode13,McQuinn15}, and the gas-bearing dwarf Antlia~B \citep[D=1.35 Mpc, $M_V$=$-$9.7 mag; ][]{Sand15,Hargis20}, both of which are members of the NGC~3109 dwarf association.  There is also KKR25, an isolated dwarf spheroidal with no signs of recent star formation or neutral gas \citep[D=1.9 Mpc, $M_V$=$-$10.9 mag;][]{Makarov12}.  A significant population of faint dwarf galaxies is expected at the edge of the Local Group \citep[e.g.][]{Tollerud18}, although recent searches (mostly looking for stellar counterparts to compact, high-velocity \hi \ clouds) have come up short, and only discovered examples of more distant, isolated dwarfs \citep{Adams13,Sand_UCHVC,Bellazzini15,Tollerud15,Tollerud16,Bennet22}.

Of particular interest are further examples of quenched dwarf galaxies at the edge of the Local Group.  This includes relatively bright objects like  Tucana ($M_V$=$-$9.5; D=890 kpc) and Cetus ($M_V$=$-$11.2; D=700 kpc), which are potential `backsplash' systems which were plausibly quenched and stripped of their gas after interacting with the Milky Way and have subsequently passed back out of the Local Group \citep[e.g.][]{Teyssier12,Buck19} -- these objects tell us about the orbital evolution of the Milky Way satellite system. Beyond the true edge of the Local Group ($\sim$2-2.5 r$_{200}$, or $\sim$750--1000 kpc) it becomes less and less likely that a dwarf galaxy has had a past interaction with the Milky Way \citep{Buck19}, and it is in this regime where true `field' dwarfs can be found.  Quenched, field dwarfs in the ultra-faint dwarf galaxy regime may cease forming stars not because of any interaction with a larger galaxy, but due to reionization \citep{Babul92,Bullock00,Benson02,Ricotti05,Jeon17,Applebaum21} or other internal mechanisms, such as supernova feedback \citep[e.g.][]{Dekel86,Maclow99}.  The discovery of such systems would provide a strong verification of galaxy formation models on small scales.

%Quenched dwarfs in the ultra-faint dwarf galaxy regime, however, may cease forming stars due to reionization or supernova feedback \citep{Applebaum21}, providing a potentially a strong test for galaxy formation models on small scales.

%There are still large regions of parameter space that have not been explored.  One in particular are faint and ultrafaint dwarfs at the edge of the LG (D$\gtrsim$ 500 kpc).  Some predictions indicate their should be many such objects, although no or few have been found.  Mention Leo P, Antlia B if possible.  These objects have retained gas and have more recent star formation.  No quenched objects, although they should exist even as backsplash systems, such as Tucana and Cetus.

Here we report the discovery of Tucana B, which to our knowledge is the first quenched, isolated ultra-faint dwarf galaxy identified in the extreme outskirts of the Local Group.  The name Tucana~B was chosen because of its constellation and the prior existence of the Tucana dwarf spheroidal; a similar naming convention has been used for other dwarfs at the edge of the Local Group (e.g. Sextans A and B; Antlia and Antlia B).  We present this new discovery in Section~\ref{sec:discovery}, and discuss follow-up optical observations in Section~\ref{sec:data}.  In Section~\ref{sec:props} we measure the basic physical properties of Tucana~B, and present an analysis of its stellar population.  In Section~\ref{sec:discussion} we place Tucana~B into context with the satellites of the Milky Way and other dwarfs in the outskirts of the Local Group.  In particular, we discuss the environment and isolation of Tucana~B, and the ramifications for reionization as a viable quenching mechanism.  We summarize and look ahead in Section~\ref{sec:summary}.

%We the X.  Found during a visual search for faint resolved dwarfs near known objects in the outskirts of the LG.

%During an initial search for faint and ultra-faint dwarf galaxy companions to the known, isolated dwarfs at the edge of the Local Group, we uncovered a new dwarf galaxy which we have dubbed Tucana~B.  

\begin{figure*}
\centering
\includegraphics[width=17.8cm]{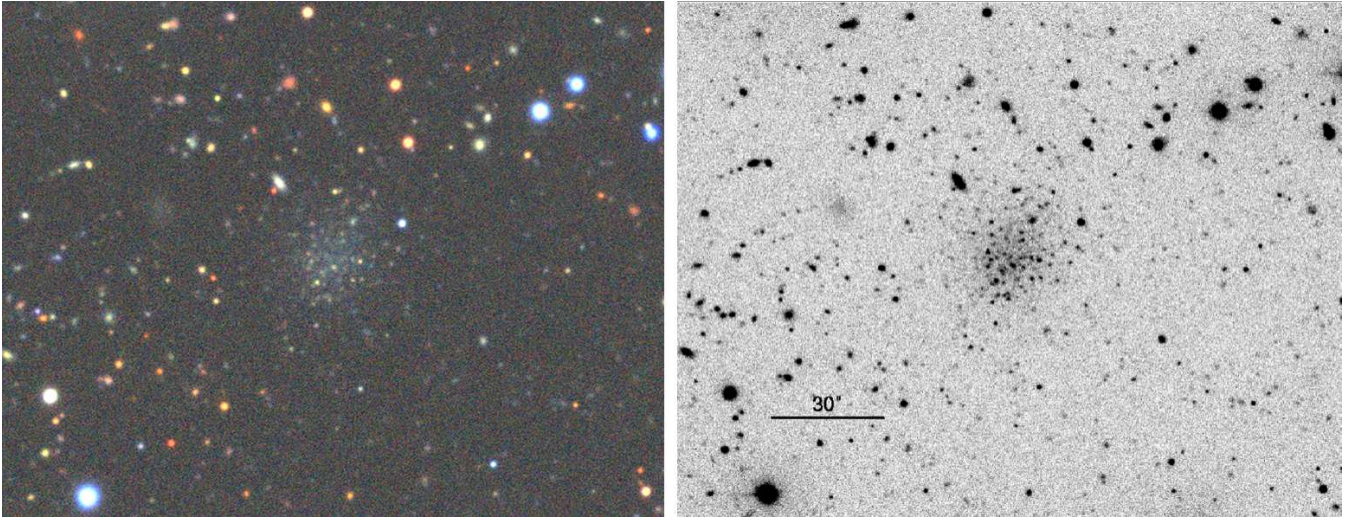}
\caption{Tucana~B as seen in the DESI Legacy Imaging Surveys sky browser (left) and in the deeper Magellan $r$-band IMACS data (right), where it is more clearly resolved into stars.  North is up and East is to the left. The diffuse object to the East of Tucana~B is likely a background dwarf galaxy associated with a galaxy group at $z$=0.036  \citep{Diaz15}, which is centered to the Northeast and has several other similar, diffuse objects associated with it. \label{fig:image}}
\end{figure*}

\section{Discovery of Tucana~B} \label{sec:discovery}

Tucana~B was found during a visual search for faint dwarf galaxy companions to the distant dwarf spheroidal galaxy Tucana \citep{Lavery92}, at $D$=890 kpc \citep{Bernard09} and $M_{V}$=$-$9.5 \citep{Savian96}. To do this, we used data from the DESI Legacy Imaging Surveys Data Release 9 \citep{Dey19} and their interactive color image viewer\footnote{\url{https://www.legacysurvey.org/viewer}}.  We uploaded a custom file to mark off a region with projected radius of 100 kpc ($\approx$6.4 deg at the distance of Tucana) and searched for visual over-densities of resolved stars with underlying diffuse light, indicative of a dwarf galaxy at the edge of the Local Group.  The field was inspected at a variety of spatial scales and contrast levels.

Tucana~B stood out during the search, and is partially resolved into stars in the Legacy Imaging Survey viewer (Figure~\ref{fig:image}).  Tucana~B is in the footprint of the Dark Energy Survey Data Release 2 \citep[DES DR2; ][]{DES_DR2}, and we downloaded photometry of the field using NOIRLab's Query Interface Tool\footnote{\url{https://datalab.noirlab.edu/query.php}}.  Tucana~B is not well-resolved in the DES $g$-band data, but has $r$ and $i$-band photometry suggestive of a resolved stellar population with an old, metal poor red giant branch (RGB).  Given this, we sought deeper ground-based optical data, which we present below.

Tucana~B is located $\sim$6 deg from the Tucana dwarf spheroidal, and is $\sim$500 kpc more distant along the line of sight (we derive a distance of D=1.4 Mpc to Tucana~B in Section~\ref{sec:dist}), and so we provisionally conclude that the two systems are not physically associated.  Further, we have performed a careful, final visual search of a 200$\times$200 kpc$^2$ region centered on Tucana~B at our inferred distance and have not identified additional resolved dwarf candidates.  We discuss the environment of Tucana~B in Section~\ref{sec:discussion}.

\section{Deep Optical Follow-Up}\label{sec:data}

%\subsection{Deep Optical Follow-up}\label{sec:deep}
Deep $g$ and $r$-band imaging was taken with the Inamori-Magellan Areal Camera \& Spectrograph \citep[IMACS;][]{IMACS} on 2021 Dec 03 (UT).  We used the $f/2$ camera, which delivers a $\sim$27.4\arcmin \ field of view and 0.2 arcsec/pixel scale.  Observations were taken in the $g$ (4$\times$300s) and $r$ (3$\times$300s) bands, with small dithers between exposures. The data were reduced in a standard way (similar to that in \citealt{Chiti20}), which included overscan subtraction and flat fielding, followed by an astrometric correction using a combination of {\sc astrometry.net} \citep{astrometry} and {\sc scamp} \citep{scamp}.  Final image stacking was accomplished with {\sc swarp} \citep{swarp} using a weighted average of the input images.  The final $g$ and $r$ band stacked images have point spread function full width half maximum values of 0.8\arcsec \ and 0.9\arcsec, respectively.  We display the final, stacked $r$-band IMACS imaging in the right panel of Figure~\ref{fig:image}.

We performed point-spread function fitting photometry on the stacked IMACS images, using {\sc daophot} and {\sc allframe} \citep{Stetson87,Stetson94}, following the general procedure described in \citet{Mutlu18}.  The photometry was calibrated to point sources in the DES DR2 catalog \citep{DES_DR2}, including a color term, and was corrected for Galactic extinction \citep{Schlafly11} on a star by star basis.  The typical color excess at the position of Tucana~B is $E(B-V)$=0.018 mag.  In the remainder of this work we present dereddened $g_0$ and $r_0$ magnitudes.

\begin{figure*}
\centering
\includegraphics[width=16.cm]{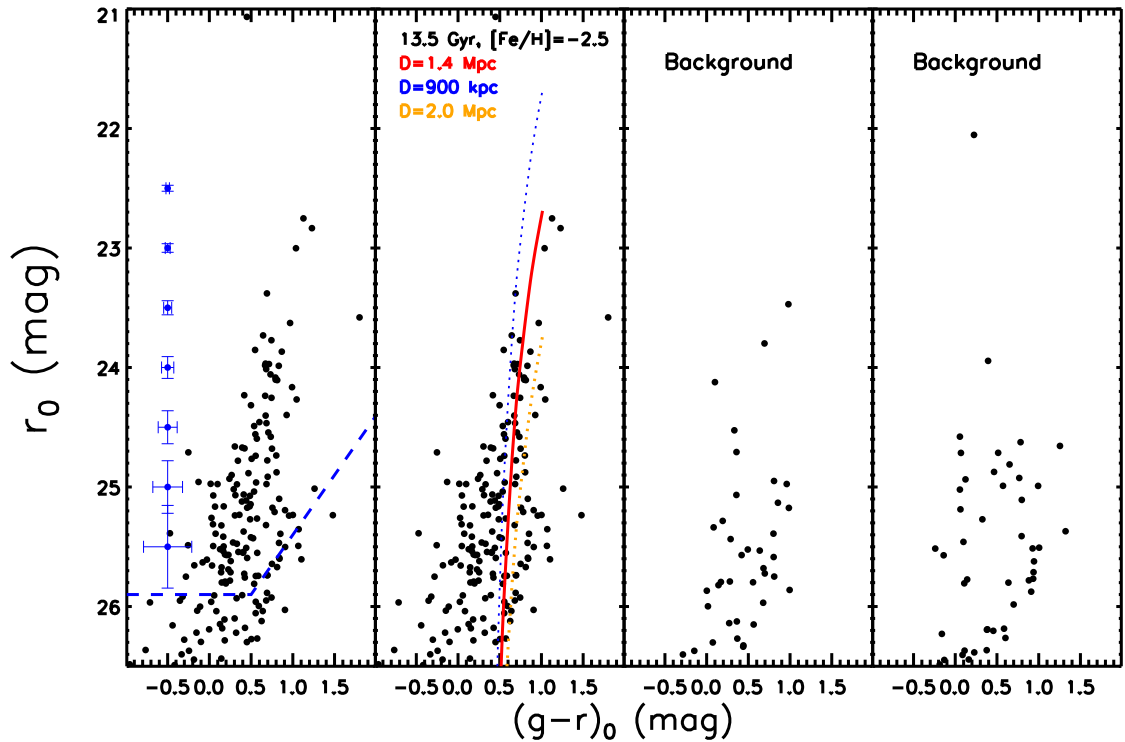}
%\includegraphics[width=8.5cm]{S190408an_globe_zoom.png}
%\includegraphics[width=\linewidth]{S190408an.png}
%\vspace{-3cm}
\caption{The color-magnitude diagram (CMD) of Tucana~B from our Magellan IMACS data (see Section~\ref{sec:data}).  In the two left panels, we display the CMD of Tucana B within 1.33 $r_h$ (16 arcsec).  Along the left side of the far-left CMD are the typical uncertainties at different $r$-band magnitudes, as determined by artificial star tests.  The dashed line marks the 50\% completeness limit.  In the center-left panel we plot 13.5 Gyr, [Fe/H]=$-$2.5 isochrones \citep{Dotter08} at different distances.  The blue dotted isochrone is at the distance of the original Tucana dwarf spheroidal (D=900 kpc), while the orange isochrone is at a distance of 2 Mpc.   The two panels on the right show randomly selected equal area background regions. There is likely significant background galaxy contamination at $r_0$$\gtrsim$ 24.5 mag.  \label{fig:CMD}}
\end{figure*}

To determine our photometric errors and completeness as a function of magnitude and color, we conduct artificial star tests with the {\sc DAOPHOT} routine {\sc ADDSTAR}, similar to previous work \citep{Sand12,Mutlu18}.  Over several iterations, we injected $\sim$10$^5$ artificial stars into our stacked images (a factor of $\sim$2 more than the number of point sources in the original image) with a range of magnitudes ($r$=18--29 mag) and colors ($g-r$=$-$0.5--1.5), and then photometered the simulated data in the same way as the original images. The 50\% (90\%) completeness level was
at $r$=25.9 (24.5) and $g$=26.4 (25.1) mag.  In Figure~\ref{fig:CMD} we show the color magnitude diagram (CMD) of Tucana B within 1.33 half-light radii $r_h$ (as derived in Section~4.3), along with several equal area “background” CMDs. We discuss the structure and stellar populations of Tucana B in the following section.

\begin{table}
\centering
    \caption{Tucana~B Properties \label{tab:props}}
     \begin{tabular}{lc}
     	\hline
     	\hline
     	Parameter & Value \\
     	\hline
        $\alpha_0$(J2000) & 22:47:00.5 $\pm$ 1.5'' \\
        $\delta_0$(J2000) & $-$58:24:27.0 $\pm$ 2.3'' \\
        $m-M$ (mag) & 25.75$^{+0.55}_{-0.45}$ \\
        Distance (Mpc) &  1.4$_{-0.3}^{+0.4}$ \\
        $M_V$ (mag) & $-$6.9$^{+0.5}_{-0.6}$\\
        $L_V$ ($L_\odot$) & $(5^{+4}_{-2})\times$10$^4$\\
        $r_h$ (arcsec) & 12$\pm$5 \\
        $r_h$ (pc) & 80$\pm$40 \\
        $\epsilon$ & $<$0.35 \\
        $\log (\mathrm{SFR_{NUV}/M_\odot \, yr^{-1}})$ & $< -5.4$ \\
        $\log (\mathrm{SFR_{FUV}/M_\odot \, yr^{-1}})$ & $< -6.0$ \\
        $\log (M_\mathrm{HI}/M_\odot)$ & $< 5.6$\\
        \hline
    \end{tabular}
\end{table}

%\footnote{\url{https://www.legacysurvey.org/viewer}}

\begin{figure}
\centering
\includegraphics[width=8.75cm]{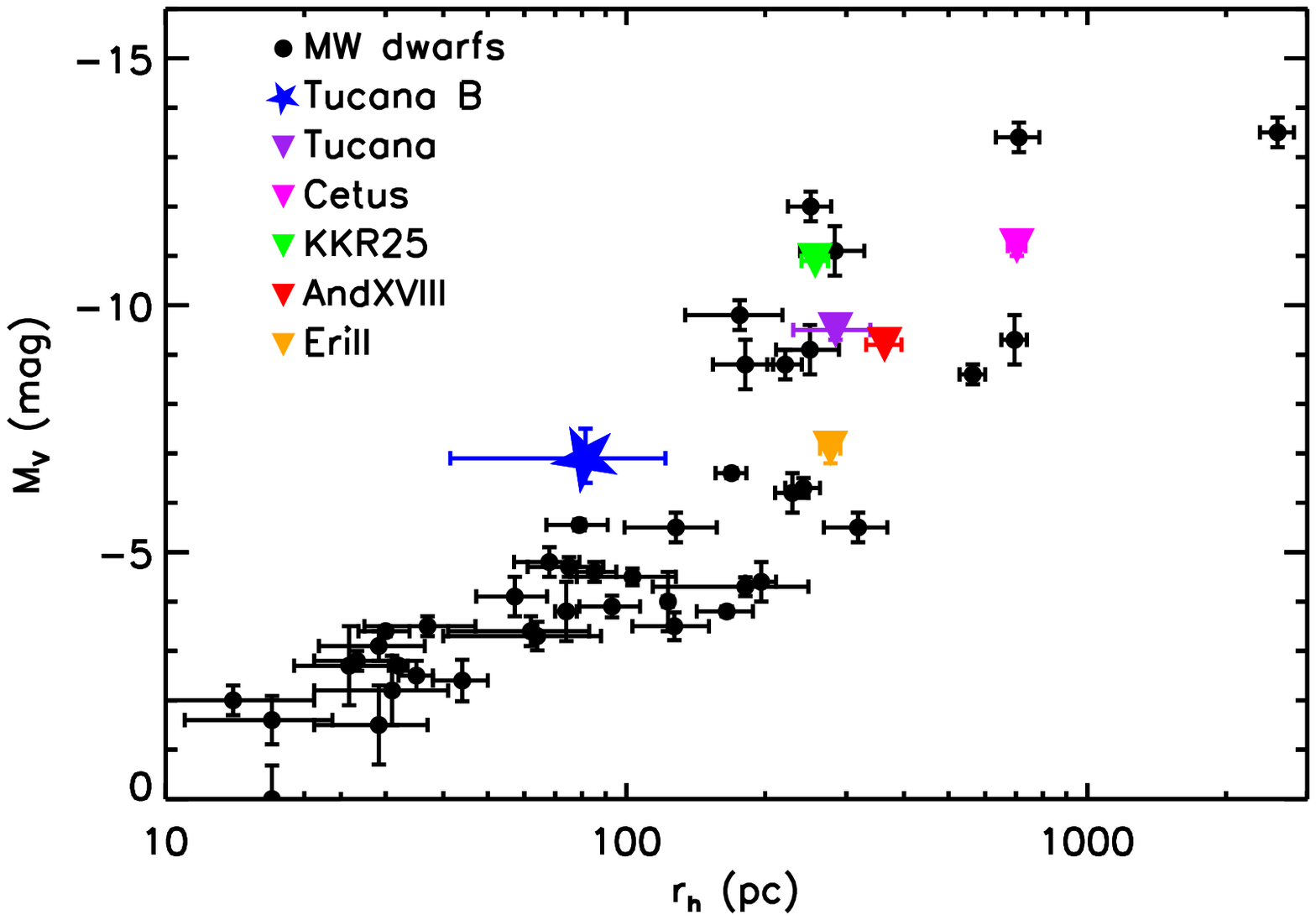}
\caption{Absolute magnitude as a function of half light radius of Tucana B and other isolated, quenched dwarfs in the outskirts of the Local Group (Cetus: \citealt{McConnachie06}; Tucana: \citealt{Savian96}; And XVIII: \citealt{McConnachie08}; KKR25: \citealt{Makarov12}; Eri II: \citealt{Crnojevic_erii}).  Also plotted are the dwarf satellites of the Milky Way \citep{Mcconachie12,Carlin17,Munoz18,Mutlu18}.  Tucana B is the faintest and smallest of the isolated, quenched Local Group dwarfs.    \label{fig:mvrh}}
\end{figure}

\begin{figure*}
\centering
\includegraphics[width=8.75cm]{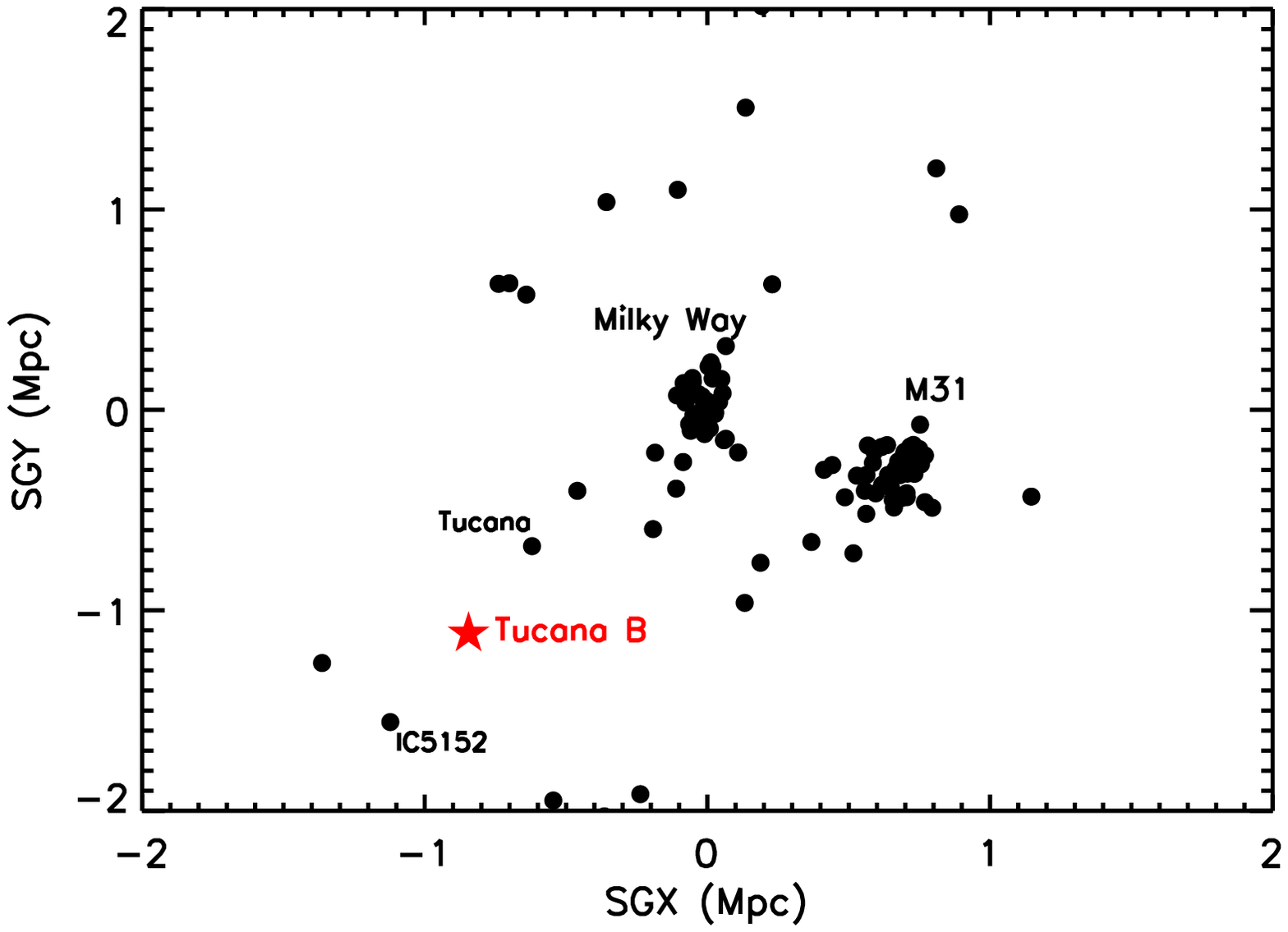}
\includegraphics[width=8.75cm]{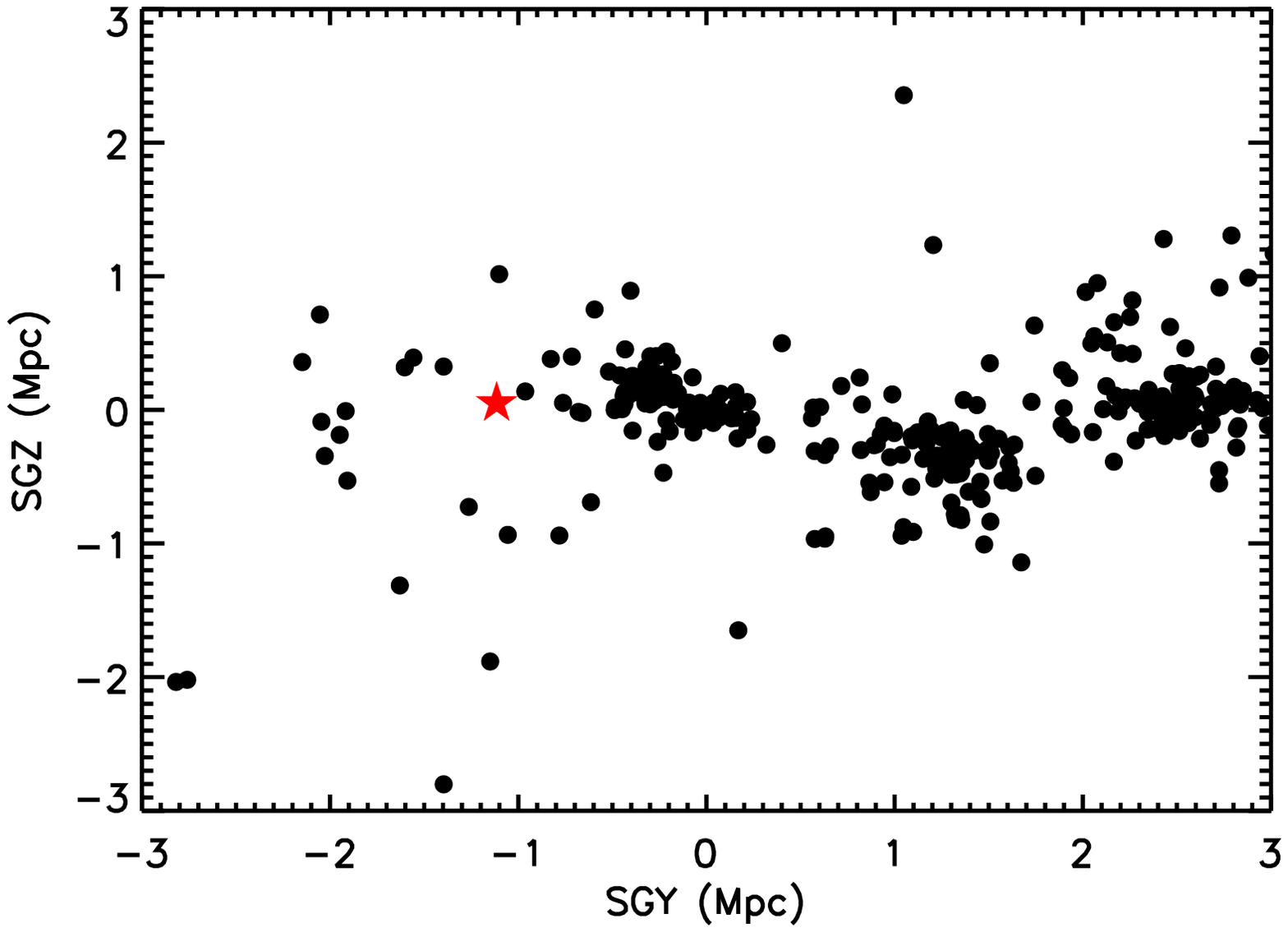}
%\vspace{-3cm}
\caption{A map of nearby galaxies (from the updated, online version of the Catalog of Neighboring Galaxies; \citealt{Kara04}) in supergalactic coordinates, highlighting the isolation of Tucana~B.  Left: In the supergalactic X-Y plane, looking down on the local plane of galaxies, Tucana~B is clearly isolated, with its nearest neighbor being the original Tucana dwarf, at a physical separation of $\sim$500 kpc. The Large Magellanic Cloud-sized galaxy IC~5152 is separated by $\sim$620 kpc (at $D$$\approx$2 Mpc; \citealt{KK02}).  These quoted separations are in 3D space,  not just in the supergalactic X-Y plane. Right: We show the supergalactic Y-Z plane, which is defined by the local plane of galaxies, used to roughly define the supergalactic coordinate system.  \label{fig:env}}
\end{figure*}

\section{Properties of Tucana~B} \label{sec:props}

In this section we measure the physical properties of Tucana~B using our new Magellan photometry, as well as archival \hi \ and {\it GALEX} UV data sets.

\subsection{Distance}\label{sec:dist}

As we discuss below, the CMD of Tucana~B seems to display a stellar population consisting of a sparsely populated old, metal poor red giant branch (Figure~\ref{fig:CMD}).
A challenge for measuring the distance to Tucana B is the lack of a well-defined tip of the red giant branch because of its intrinsic faintness, where few or no stars populate the upper regions of the red giant branch, as has been discussed in several previous works \citep[e.g.][]{Madore95,WeiszHB19,Carlin21,Mutlu22}. The ground-based data are also too shallow to identify a horizontal branch or any associated RR Lyrae stars, both of which could provide distance estimates. 

Instead of using a tip of the red giant branch derived distance, we measure the distance to Tucana B using a CMD-fitting technique, comparing the number of stars consistent with several old, metal poor theoretical isochrones, and adopting a methodology similar to that used for several of the Milky Way ultra-faint dwarfs \citep[e.g.][]{Walsh08,Sand09}.  We use the Dartmouth isochrones \citep{Dotter08}, focusing on tracks with stellar ages of 13.5 Gyr and low metallicities ([Fe/H] of $-$2.5, $-$2.0 and $-$1.5).  We include all Tucana~B stars with $r_0$$<$25.5 mag within 0.35\arcmin \ of its center, which visually encapsulates the bulk of its main body. Each isochrone fiducial is shifted through 0.025 mag intervals in distance modulus ($m-M$) from 24.5 to 27.0 mag ($\sim$0.8--2.5 Mpc).  At each of these steps, the number of stars consistent with the fiducial is tabulated.  The selection region for stars to be counted is simply defined by two red/blue boundaries in $(g-r)_0$ color for a given $r_0$ magnitude, as determined by the uncertainties found in our artificial star tests.  Background stars are also accounted for by running the identical procedure over an appropriately scaled background region, and then subtracting this number from that of the Tucana B selection.  Implicit in this technique is the assumption that Tucana B has a single age, exclusively old ($\sim$13.5 Gyr) stellar population.  We have not explored other isochrone model tracks from different groups, or at different ages, which may expand the allowed distance range discussed below.  We also assume no internal extinction associated with Tucana~B, and only include a Milky Way component. 

The best-fit distance moduli for the 13.5 Gyr, [Fe/H]=$-$2.0 and $-$2.5 isochrones are $m-M$=25.6 and 25.9 mag ($D$=1.3 and 1.5 Mpc), respectively.  We found the best-fit distance for the 13.5 Gyr, [Fe/H]=$-$1.5 isochrone unsatisfactory, implying a distance too nearby and too red to properly match the data (see discussion in the next paragraph).  Given the two satisfactory matches for a very metal-poor population, we take the mean of those two measurements as our distance modulus ($m-M$=25.75; $D$=1.4 Mpc), and continue to discuss our uncertainties below.

%, and the range as the uncertainty (see Table~\ref{tab:props}).  It is possible that this underestimates the true uncertainty in our measurement; we recommend deeper, high spatial resolution space-based data to measure a higher precision distance to Tucana~B, and discuss plausible alternative distances next (which we ultimately rule out).

Given the projected proximity of Tucana~B to the original Tucana dwarf spheroidal galaxy, which is at $D$=900 kpc, we overplot an old metal-poor isochrone (13.5 Gyr, [Fe/H]=$-$2.5) at that distance onto the CMD in Figure~\ref{fig:CMD}.  From this, it is clear that Tucana~B is at a larger distance than the original Tucana dwarf, otherwise it would imply that the upper 1--2 magnitudes of the red giant branch in Tucana~B were completely unpopulated. Thus, the two objects are not likely physically associated, and the lower distance limit to Tucana~B is greater than 900 kpc.

As another point of discussion, inspection of the CMD shows three stars at $r_0$$\approx$22.7--23.0 and $(g-r)_0$$\approx$1.1; the membership status and provenance of these stars is important for assessing the distance of Tucana~B.  First, aside from our baseline assumption that these stars are upper red giant branch members, these stars may be asymptotic giant branch (AGB) stars associated with Tucana B itself, which would imply that the brightest red giant branch stars are at $r_0$$\approx$23.5 mag.  Alternatively, these stars could be foreground stars, although there are very few contaminant stars at that position in color-magnitude space (see right panels of Figure~\ref{fig:CMD}).  In either, it may imply a larger distance to Tucana~B, although we discount this possibility because it does not match the data.  We over plot the same 13.5 Gyr, [Fe/H]=$-$2.5 isochrone, but at a distance of $D$=2.0 Mpc, corresponding to a scenario where the three stars at $r_0$$\approx$22.7--23.0 are either foreground contaminants or AGB stars.  While this isochrone roughly matches the putative magnitude of the tip of the red giant branch in this scenario, it is skewed redward of the main locus of Tucana~B stars, and cannot be pushed further blueward as it would imply an unrealistically low metallicity for Tucana~B. For this reason, Tucana~B must be at a closer distance than 2 Mpc, and the three bright aforementioned stars are not likely to be AGB stars or exclusively foreground contaminants.  

To determine a distance uncertainty to Tucana~B more concrete than the broad limits discussed above, we carefully moved old, metal poor isochrones (both 13.5 Gyr, [Fe/H]=$-$2 and $-$2.5) by eye through different distances around our best fit of $m-M$=25.75 ($D$=1.4 Mpc), and noted where clear departures between isochrone and data occur.  For distances much closer than $m-M$=25.75, the isochrones would imply a completely unpopulated upper red giant branch, and the three stars at $r_0$$\approx$22.7--23.0 become too red to match a metal-poor isochrone. From this inspection we estimate a lower distance limit of $m-M$=25.3 ($D$=1.1 Mpc).  For distances farther than $m-M$=25.75, we were most concerned about the isochrone skewing redwards of the data, without much concern for the three stars at $r_0$$\approx$22.7--23.0 as these may be contaminants or member AGB stars in this scenario, as discussed above.  This leads to a visual upper distance limit of $m-M$=26.3 ($D$=1.8 Mpc).  While we emphasize that space-based data down to the horizontal branch is necessary for a definitive distance to Tucana~B, the ground-based CMD indicates a $m-M$=25.75$^{+0.55}_{-0.45}$ mag ($D$=1.4$_{-0.3}^{+0.4}$ Mpc).

\subsection{Stellar Population}\label{sec:pop}

Based on the CMD and our distance constraints, Tucana~B appears to only consist of an old metal-poor stellar population (Figure~\ref{fig:CMD}).  Other recently discovered dwarfs at the edge of the Local Group, such as Leo~P ($M_V=-9.3$) and Antlia~B ($M_V=-9.7$), have clear signs of star formation in the form of blue main sequence stars \citep{McQuinn15,Hargis20}, but any plausible blue member stars in Tucana~B are small in number and consistent with background contamination (Figure~\ref{fig:CMD}).  Its CMD is most analogous to the ultra-faint dwarf galaxies of the Milky Way, which only consist of an old stellar population, with the exception of Leo T \citep{dejong08,Weisz12}.  %It is important that Tucana~B be observed to deeper depths, most likely with {\it HST} in order to further constrain its stellar population.

To further assess any possible young stellar population, we search for coincident UV emission with data from the  {\it Galaxy Evolution Explorer} ({\it GALEX}; \citealt{galex}) archive, which is sensitive to star formation on $\lesssim$100 Myr time scales \citep[e.g.][]{Lee11}.  We adopt the methodology of \citet{Karunakaran21}, using a 16 arcsec aperture (1.33$r_h$), finding no detection at the position of Tucana~B.  We assess our detection limits by placing 1000 random apertures over the {\it GALEX} field (after masking bright objects).  We then translate these NUV and FUV flux limits to star formation rate limits using the relations of \citet{Iglesias-Paramo+2006}, finding $\log (\mathrm{SFR_{NUV}/M_\odot \, yr^{-1}}) < -5.4$  and $\log (\mathrm{SFR_{FUV}/M_\odot \, yr^{-1}}) < -6.0$, respectively.  Both limits are more than an order of magnitude more stringent than nearly all UV detections in satellite galaxies around Milky Way-like halos \citep[e.g.][]{Karunakaran21}, again emphasizing the lack of star formation in Tucana~B.

It is important to note that the current data set cannot rule out the presence of a faint intermediate age stellar population, greater than $\sim$500 Myr old.  Such intermediate age stellar populations do exist in faint dwarf galaxies in the Local Group, for instance in the {\it HST} imaging of Andromeda XVI \citep{Weisz14M31}.  To be sensitive to such stellar populations will require deep space-based observations, as we continue to emphasize in this work.

\subsection{Stellar Structure and Luminosity}\label{sec:struct}

Tucana B is a distant stellar system with only $\sim$100 resolved stars in our Magellan/IMACS data.  As in previous work on similarly sparse systems \citep[e.g.][]{Sand14}, we fit an exponential profile to the two-dimensional distribution of stars consistent with the red giant branch of Tucana B using the maximum likelihood technique of \citet{Martin08}.  We select stars for this analysis which are consistent with the best-fitting Dartmouth isochrone, at a distance of $D=$1.4 Mpc as found in Section~\ref{sec:dist}, after taking into account the photometric uncertainties for stars brighter than the 50\% completeness limit.  The fit includes the central position ($\alpha_0$,$\delta_0$), position angle ($\theta$), ellipticity ($\epsilon$), half-light radius ($r_h$), and a constant background surface density as free parameters.  Uncertainties on each parameter were calculated through a bootstrap resampling analysis, with 1000 iterations.  As a check on our results, we repeated the calculations while only including red giant branch stars down to $r_0$=25.4 mag, a half magnitude brighter than our initial iteration; the derived structural parameters did not change to within the uncertainties.

The results of the structural analysis are shown in Table~\ref{tab:props}.  Tucana~B has a half light radius of 80$\pm$40 pc (this includes our distance uncertainty). Its ellipticity is not well-constrained: we find an upper limit of $<$0.35 at 95\% confidence, reinforcing the roughly circular shape of the dwarf seen in Figure~\ref{fig:image}.  Given this, no constraint on the position angle is possible.  The central position of the new dwarf is constrained to a couple of arcseconds. % Further constraints on the structure of Tucana~B will require deeper photometry, possibly from space, in order to resolve more stars.

To derive the luminosity of Tucana~B we employ the methodology of \citet{Martin08}, which is appropriate for faint dwarf galaxies in the `CMD shot noise' regime, where the presence or absence of individual stars in the upper RGB can greatly influence the overall luminosity.  First, we build a well-populated CMD by using a [Fe/H]=$-$2, 13.5 Gyr isochrone and a Salpeter initial mass function.  We convolve this with our measured completeness and photometric uncertainties, shifted to $D=1.4$ Mpc.  We then randomly selected the same number of stars from the simulated CMD as was found in our maximum likelihood analysis, taking into account the luminosity in the simulated population below our detection limit.  We repeat this process 100 times, taking the median and standard deviation as our final absolute magnitude and uncertainty (we also add the distance uncertainty in quadrature for our final absolute magnitude uncertainty).  We convert to the $V$-band using the filter transformation equations of \citet{Jordi06}, and ultimately find $M_V$=$-$6.9$^{+0.5}_{-0.6}$ mag  ($L_V$=$(5^{+4}_{-2})\times$10$^4$ $L_{\odot}$).  Based on this, Tucana~B is a true ultra-faint dwarf galaxy, similar in luminosity to Eridanus~II \citep{Crnojevic_erii}.

\subsection{Gas Content}\label{sec:gas}

Given the isolation and apparent lack of recent star formation in Tucana~B, it is important to assess its neutral gas content.
The deepest available \hi \ observations in the direction of Tucana~B are from the Galactic All Sky Survey \citep[GASS;][]{McClure-Griffiths2009,Kalberla2010,Kalberla2015}, with the caveat that these observations only extend to a redshift of $\sim$500~\kms. Tucana~B is located in a field surrounded by complex foreground \hi \ features associated with the Milky Way and Magellanic Clouds \citep[e.g.][]{Westmeier2018}. This complicates the search for an \hi \ counterpart of Tucana~B as, without an a priori redshift, any \hi \ emission along the line-of-sight could be associated with the Milky Way.  Although difficult, it will be important to measure a stellar velocity for Tucana~B for this task.

In the GASS data there is a clump of \hi \ emission that peaks approximately 20\arcmin \ to the SW of Tucana~B (note that the spatial resolution of the GASS data is 16\arcmin). This clump (at $cz_\odot \approx 220$~\kms) forms a distinct, almost point-like (at the resolution of GASS) feature, but is surrounded by Milky Way emission. Therefore, it is highly likely that this feature is merely associated with the Milky Way and not Tucana~B, but we cannot robustly exclude either option with the available data.

There is no other candidate feature in the GASS data, therefore, if we assume this feature is not associated with Tucana~B then we can proceed to set an upper limit on its \hi \ mass. The typical rms noise of GASS is 53~mK in 0.82~\kms \ channels. Approximating the Parkes radio telescope gain as 0.7~K/Jy gives a 3$\sigma$ sensitivity of $\log (M_\mathrm{HI}/M_\odot) < 5.6$ for a source of 20~\kms \ velocity width and a distance of 1.4~Mpc.  Future high-resolution, interferometric observations, potentially with MeerKAT or the Australia Telescope Compact Array (ATCA) are necessary to further constrain the HI content of Tucana~B.% For reference, Leo~T has a HI gas mass of $\log (M_\mathrm{HI}/M_\odot) = 5.4$, indicating that the Tucana~B limit is not very stringent.

\section{Discussion} \label{sec:discussion}

%To put Tucana~B in context, we first examine its environment.  

%compare physical properties with objects in the LG and its outskirts.  

Tucana B is a unique discovery for a dwarf galaxy just beyond the Local Group, given both its luminosity, isolation and its lack of apparent star formation or neutral gas.  Here we discuss the physical properties and environment of Tucana~B in the broader context of faint dwarf galaxies as a population.%; we also discuss, as well as our understanding of galaxy formation and evolution.

\subsection{Local Group and Isolated Dwarf Comparisons}

In Figure~\ref{fig:mvrh} we plot a size -- luminosity relation featuring the satellites of the Milky Way, along with quenched dwarfs in the outskirts of the Local Group.  We also highlight  Eridanus II, which is situated near the virial radius of the Milky Way itself, but is devoid of recent star formation and gas.

In comparison to the Milky Way satellites, Tucana~B has a luminosity similar to Eridanus~II \citep{Crnojevic_erii,Simon21}, and is slightly more luminous than the ultra-faint dwarfs Bo{\"o}tes~I \citep[$M_V$=$-$6.0 mag;][]{Munoz18} and Hercules \citep[$M_V$=$-$6.2 mag;][]{Sand09}.  Interestingly, Tucana~B is significantly more compact than either of these systems, which have half light radii $\gtrsim$200 pc.  Its half light radius is comparable to much fainter  satellites such as Canes Venatici~II ($M_V$$\approx$$-$4.6 mag; $r_h$$\approx$85 pc; \citealt{Sand12}).

Despite its compact nature for its luminosity, Tucana~B is only a mild outlier in comparison to the Milky Way dwarf population as a whole, especially considering its large uncertainties.  If a larger population of field ultra-faint dwarfs display a more compact stellar distribution than their counterparts near the Milky Way, it may point to dwarf mergers, tidal encounters or similar mechanisms puffing up the stellar distribution in the Milky Way sample \citep[e.g.][]{Frings17,Chiti21,Tarumi21}, although this is just speculation at this point.

\subsection{Environment} \label{sec:env}

To investigate the environment of Tucana~B, we plot the known dwarfs and galaxies of the Local Volume in two projections of the supergalactic coordinate system in Figure~\ref{fig:env}.  We focus on the supergalactic X-Y projection, as the Y-Z plot simply shows the local plane of galaxies used to roughly define the supergalactic coordinate system; note that Tucana~B is on this plane.  Given our search area, it is no surprise that the nearest galaxy to Tucana~B is the original Tucana dwarf, at a physical separation of $\sim$500 kpc, followed by the low-mass galaxy IC5152 ($\sim$620 kpc separation).  Both of these separations are beyond the virial radius of either low-mass galaxy (even with the considerable Tucana~B distance uncertainties). As mentioned elsewhere, it will be important to measure a velocity to Tucana~B to further assess whether it interacted with either low mass system in the past.  Tucana~B is also well beyond the virial radius of the Milky Way itself, which is  between $r_{vir}$=200--300 kpc \citep[e.g.][]{Klypin02,Putnam21}, putting it at $\gtrsim$4.5 $r_{vir}$.  Barring future discoveries, Tucana~B is one of the most isolated galaxies within $\sim$2 Mpc.

%, and barring future discoveries, Tucana~B is one of the most isolated galaxies within $\sim$2 Mpc.  To be clear, Tucana~B is also well beyond the virial radius of the Milky Way itself, which is 

\subsection{The Nature of Tucana~B} \label{sec:nature}

Tucana~B is an apparently quenched, isolated ultra-faint dwarf galaxy which may have deep implications for our view of low mass galaxy formation, depending on its origins.  

First, it is possible that Tucana~B is a so-called backsplash galaxy, and may have had a previous encounter with the Milky Way, quenching its star formation and stripping it of its gas before being ejected out to large galactocentric radii.  Several systems in the outskirts of the Local Group have been identified as possible backsplash systems \citep[see, for instance, Table 4 in ][for one sample]{Buck19}.  In order to fully assess this scenario, a velocity measurement of Tucana~B is necessary, and both a stellar velocity measurement and deeper \hi \ observations should be prioritized.  Nonetheless, simulations suggest it is very unlikely that Tucana~B is a backsplash system, as most such galaxies should  be at $\lesssim$2.5 $r_{vir}$ \citep[e.g.][]{Teyssier12,Diemer15,Buck19,Applebaum21}, which is significantly closer to the Milky Way than Tucana~B.%well within its current distance from the Milky Way.

%Applebaum21 note that most galaxies beyond 1.5r_vir are `field galaxies' while most dwarfs around the virial radius are backsplash systems.
%Diemer+15 show that backsplash galaxies can be found within 2.5rvir.

%Is Tucana~B a splashback galaxy or an isolated field dwarf with no previous interaction.  It will be important to measure a optical velocity, and to perform deeper HI observations. as this will address quenchign and any past interaction with the MW.  Without these measurements, we can still 

It is also possible that Tucana~B is a faint example of a transition dwarf galaxy, which exhibit HI emission, but show no signs of recent star formation \citep{Skillman03,Weisz11}, possibly because they are in between star formation episodes \citep[see e.g.][]{elbadry16}.  While no HI emission is detected in Tucana~B, more stringent limits are necessary to rule out this scenario (preferably down to M$_{HI}$/L$_V$$\approx$1; e.g. \citealt{Putnam21}), along with deeper photometry to probe intermediate age star formation events.

As the backsplash scenario is unlikely, it points to Tucana~B being a true quenched field ultra-faint dwarf galaxy (again with the caveat that deeper optical and \hi \ data are necessary).  Without any interaction with the hot halo or gravitational field of a massive galaxy, reionization and/or some other internal mechanism (i.e. supernova feedback) is likely responsible for the gas free and quenched status of Tucana~B.  Such a scenario is seen in recent simulations of field dwarf galaxies \citep[e.g.][]{Jeon17,Rey20,Applebaum21}, although even here it is possible for faint dwarf galaxies to be quenched by other mechanisms, such as ram pressure stripping by gas in the cosmic web itself \citep[e.g.][]{BL13}.

It has long been recognized that reionization can essentially `boil' the gas out of the dark matter potential wells of ultra-faint dwarf galaxies, explaining their dearth of stars and the possibility that some dark matter halos never form stars at all \citep[e.g.][]{Babul92,Bullock00,Benson02,Ricotti05}.  Observations of the ultra-faint dwarfs around the Milky Way lend support to this picture, as {\it HST} observations down to the oldest main sequence turnoff reveal nearly synchronous star formation at very early times ($\sim$13 Gyr ago), with little to no star formation since the reionization epoch \citep{Brown14,Weisz14}.  However, as the Milky Way ultra-faint dwarfs have also experienced its hot halo and significant tidal forces, it is difficult to distinguish between all of these mechanisms as the primary source for their quenched, gas-free status.  Thus, Tucana~B may provide strong confirmation of reionization's role in influencing galaxy formation and evolution at the lowest mass scales.

\section{Summary \& Future Outlook}\label{sec:summary}

We have presented the discovery of Tucana~B, an ultra-faint dwarf galaxy in the extreme outskirts of the Local Group ($D$$\approx$1.4 Mpc).  Its luminosity ($M_V$=$-$6.9 mag) and apparent lack of star formation and neutral gas make it unique among recent discoveries  at this approximate distance \citep[e.g.][]{Makarov12,Giovanelli13,Sand15}.  The isolation of Tucana~B also suggests that its star formation was quenched by reionization or some other internal mechanism, rather than interaction with a larger galaxy halo, although further data is necessary to solidify these results. Even if it is found that Tucana~B has an intermediate age stellar population, or contained a reservoir of HI gas, it will still be a novel ultra-faint dwarf with properties distinct from those found around the Milky Way. It is likely that similar systems to Tucana~B are waiting to be discovered, although its semi-resolved status in the discovery DECam data point to the need for a tailored search.  A complete census of such objects is necessary to understand the demographics of the field ultra-faint dwarf galaxy population.

Tucana B is a prime target for future space-based follow-up to pin down its structure and star formation history, possibly down to the oldest main sequence turnoff.  In particular, Tucana~B may provide a definitive opportunity to understand the role that reionization plays in the quenching of the faintest galaxies.  As discussed, {\it HST} observations of the Milky Way's ultra-faint dwarfs down to the oldest main sequence turnoff reveal early, nearly synchronous star formation \citep{Brown14,Weisz14}.  There are hints that those ultra-faint dwarfs associated with the Magellanic Clouds had a slightly different star formation epoch than those of the Milky Way \citep{Sacchi21}, and probing further systems and environments is essential for confirming and extending these results. Prior to the advent of the {\it James Webb Space Telescope}, similar observations beyond the Local Group were prohibitive, but the Near Infrared Camera (NIRCam) enables such studies out to $\sim$2 Mpc \citep[e.g.][]{Weisz19}.  Tucana~B, and similar systems to be discovered in the near-future \citep[e.g. see predictions and simulated data in][ respectively]{Rodriguez19,Mutlu_sims}, will be a crucial data set for understanding reionization's effect on star formation and subsequent cessation in the smallest dark matter halos.

\acknowledgments

DJS acknowledges support from NSF grants AST-1821967 and 1813708. BMP is supported by an NSF Astronomy and Astrophysics Postdoctoral Fellowship under award AST-2001663. Research by DC is supported by NSF grant AST-1814208. AK acknowledges financial support from the State Agency for Research of the Spanish Ministry of Science, Innovation and Universities through the ``Center of Excellence Severo Ochoa" awarded to the Instituto de Astrof\'{i}sica de Andaluc\'{i}a (SEV-2017-0709) and through the grant POSTDOC\_21\_00845 financed from the budgetary program 54a Scientific Research and Innovation of the Economic Transformation, Industry, Knowledge and Universities Council of the Regional Government of Andalusia.
F.W. thanks the support provided by NASA through the NASA Hubble Fellowship grant \#HF2-51448 awarded by the Space Telescope Science Institute, which is operated by the Association of Universities for Research in Astronomy, Incorporated, under NASA contract NAS5-26555.
AC is supported by a Brinson Prize Fellowship at UChicago/KICP. KS acknowledges support from the Natural Sciences and Engineering Research Council of Canada (NSERC).

This research uses services or data provided by the Astro Data Lab at NSF's National Optical-Infrared Astronomy Research Laboratory. NOIRLab is operated by the Association of Universities for Research in Astronomy (AURA), Inc. under a cooperative agreement with the National Science Foundation.

The work used images from the Dark Energy Camera
Legacy Survey (DECaLS; Proposal ID 2014B-0404; PIs:
David Schlegel and Arjun Dey). Full acknowledgment
at \url{https://www.legacysurvey.org/acknowledgment/}. 

This research is based on observations made with the {\it Galaxy Evolution Explorer}, obtained from the MAST data archive at the Space Telescope Science Institute, which is operated by the Association of Universities for Research in Astronomy, Inc., under NASA contract NAS 5–26555.

This project used public archival data from the Dark Energy Survey (DES). Funding for the DES Projects has been provided by the U.S. Department of Energy, the U.S. National Science Foundation, the Ministry of Science and Education of Spain, the Science and Technology Facilities Council of the United Kingdom, the Higher Education Funding Council for England, the National Center for Supercomputing Applications at the University of Illinois at Urbana-Champaign, the Kavli Institute of Cosmological Physics at the University of Chicago, the Center for Cosmology and Astro-Particle Physics at the Ohio State University, the Mitchell Institute for Fundamental Physics and Astronomy at Texas A\&M University, Financiadora de Estudos e Projetos, Funda{\c c}{\~a}o Carlos Chagas Filho de Amparo {\`a} Pesquisa do Estado do Rio de Janeiro, Conselho Nacional de Desenvolvimento Cient{\'i}fico e Tecnol{\'o}gico and the Minist{\'e}rio da Ci{\^e}ncia, Tecnologia e Inova{\c c}{\~a}o, the Deutsche Forschungsgemeinschaft, and the Collaborating Institutions in the Dark Energy Survey.
The Collaborating Institutions are Argonne National Laboratory, the University of California at Santa Cruz, the University of Cambridge, Centro de Investigaciones Energ{\'e}ticas, Medioambientales y Tecnol{\'o}gicas-Madrid, the University of Chicago, University College London, the DES-Brazil Consortium, the University of Edinburgh, the Eidgen{\"o}ssische Technische Hochschule (ETH) Z{\"u}rich,  Fermi National Accelerator Laboratory, the University of Illinois at Urbana-Champaign, the Institut de Ci{\`e}ncies de l'Espai (IEEC/CSIC), the Institut de F{\'i}sica d'Altes Energies, Lawrence Berkeley National Laboratory, the Ludwig-Maximilians Universit{\"a}t M{\"u}nchen and the associated Excellence Cluster Universe, the University of Michigan, the National Optical Astronomy Observatory, the University of Nottingham, The Ohio State University, the OzDES Membership Consortium, the University of Pennsylvania, the University of Portsmouth, SLAC National Accelerator Laboratory, Stanford University, the University of Sussex, and Texas A\&M University.
Based in part on observations at Cerro Tololo Inter-American Observatory, National Optical Astronomy Observatory, which is operated by the Association of Universities for Research in Astronomy (AURA) under a cooperative agreement with the National Science Foundation.

%Local Universe research by D.J.S. is supported by NSF grants 

% Time domain research by D.J.S.\ is also supported by NSF grants AST-1821987, 1813466, 1908972, \& 2108032, and by the Heising-Simons Foundation under grant \#2020-1864. 

%We acknowledge ESA Gaia, DPAC and the Photometric Science Alerts Team\footnote{ \url{http://gsaweb.ast.cam.ac.uk/alerts}}.

\vspace{5mm}
\facilities{Magellan:Baade (IMACS), {\it GALEX}, Parkes, Blanco}

\software{  astropy \citep{2013A&A...558A..33A,astropy}, {\sc astrometry.net} \citep{astrometry}
The IDL Astronomy User's Library \citep{IDLforever}, DAOPHOT \citep{Stetson87,Stetson94}, {\sc scamp} \citep{scamp}, {\sc swarp} \citep{swarp}
          }

\bibliography{biblio}
\bibliographystyle{aasjournal}

\end{document}